# Communicating astrobiology and the search for life elsewhere: speculations and promises of a developing scientific field in newspapers, press releases and papers

Danilo Albergaria[1,2] *, Pedro Russo[1,2], Ionica Smeets[2], Thilina Heenatigala[3] and Dallyce Vetter[2]

[1] Department of Science Communication & Society and Leiden Observatory, Leiden University, Leiden, the Netherlands

[2] Department of Science Communication & Society, Institute of Biology, Leiden University, Leiden, the Netherlands

[3] Earth-Life Science Institute (ELSI), Institute of Science Tokyo, Tokyo, Japan

*Corresponding authorE-mail: albergaria@strw.leidenuniv.nl (DA)

Short title: **Communicating astrobiology: speculations and promises of a developing scientific field**

# Abstract:

This study examines the communication of astrobiology and the Search for Life Elsewhere (SLE) in academic papers, press releases, and news articles over three decades. Through a quantitative content analysis, it investigates the prevalence of speculations and promises/expectations in these sources, aiming to understand how research results are portrayed and their potential impact on public perception and future research directions. Findings reveal that speculations and promises/expectations are more frequent in news articles and press releases compared to academic papers. Speculations about conditions for life and the existence of life beyond Earth are common, particularly in news articles covering exoplanet research, while promises of life detection are rare. Press releases tend to emphasize the significance of research findings and the progress of the field. Speculations and promises/expectations in news articles often occur without attribution to scientists and in quotes of authors of the studies, and slightly less so in quotes of outside experts. The study highlights the complex dynamics of science communication in astrobiology, where speculations and promises can generate public excitement and influence research funding, but also risk misrepresenting scientific uncertainty and creating unrealistic expectations. It underscores the need for responsible communication practices that acknowledge the speculative dimension of the field while fostering public engagement and informed decision-making.

# 1. Introduction

Astrobiology is a broad, inter-multidisciplinary field focused on answering fundamental questions about life as a possible universal phenomenon: how did it emerge, what are the conditions for its existence and what is its distribution in the universe? [1, 2, 3]. The branch dedicated to the third question, here called the Search for Life Elsewhere (SLE), touches upon a theme of enduring pervasiveness in modern popular culture [4]. For most of history, whether life existed beyond Earth was a matter intractable to empirical and observational inquiry. However, in the late 1950's and early 1960's, this started to change with the inception of the exobiology research programme of the National Aeronautics and Space Administration (NASA) in the United States of America (USA) and the first efforts in the Search for Extraterrestrial Intelligence (SETI). This was followed by a string of relevant discoveries in the 1970's through to the 1990's: extremophiles in hydrothermal vents, organic molecules in molecular clouds, planet-forming disks around young stars, and the first detections of exoplanets [5].

In 1998, the term astrobiology came to the forefront in the public sphere when the NASA Astrobiology Institute (NAI) was established. A rapid process of international institutionalization ensued: in 1999, the Centro de Astrobiologia was created in Spain, and a year later, the Cardiff Centre for Astrobiology was founded (but closed in 2011), culminating in the European Astrobiology Network Association (EANA) in 2001 and the constitution of similar institutions in other regions. In Latin America, the Sociedad Mexicana de Astrobiología was constituted in 2002 in Mexico, while the Laboratório de Astrobiologia - AstroLab, of the University of São Paulo, was inaugurated in Brazil in 2012. In Asia, the Indian Astrobiology Research Foundation (IARF) was established in 2006, while the Earth-Life Science Institute (ELSI) was founded in Tokyo, Japan, in 2012. This process of institutionalization reflected a gain of momentum for the field, in a period in which it drifted into mainstream science and stirred optimism about the prospects of success in finding life beyond Earth.

## 1.1. Astrobiology in the public sphere

Although public interest in astrobiology is usually assumed to be widespread [6 - 9], it is still a poorly understood phenomenon, with evidence indicating that its appeal is far from universal in the USA [10]. Given that the foundation of NAI in 1998 is known to be associated with a peak in public interest generated by a claim of discovery of fossilized martian life in a meteorite [11, 12] (more on that below), the institutionalization of astrobiology in many different countries suggests the field touches on subjects of a corresponding level of public interest. Although research points towards regional variations in public interest and support for science – Europeans have very positive views [13], while public trust is declining in the USA [14] and Brazil shows a relatively low public awareness of science [15] -, public perceptions of science inform attitudes towards public funding for research (i.e. those who have a positive perception of science tend to be favorable to public funding of research) [16] and variation in public support may ultimately impact funding decisions [17], which partially depend on the effective communication of research results [18]. This underlines the importance of accurately communicating a realistic view of the current scientific knowledge about the possible existence of life elsewhere in the universe, the conditions for its existence, and the prospects of finding it.

During the last three decades, there have been a few high-profile cases in which claims of possible discoveries and scientific breakthroughs related to astrobiology and the SLE have made the headlines in many international newspapers (see S5 Figure for a visual representation of these cases in the historical context of astrobiology). In August of 1996, a study published in the journal *Science* by a team of various institutions in the USA and Canada, including NASA, claimed to have obtained "evidence for primitive life in early Mars" in combined features resembling microfossils in the ALH84001 Mars meteorite [19]. This publication prompted a public announcement by then USA president Bill Clinton reflecting on the significance of the possible discovery and a NASA press conference with scientists involved in the study, which ignited a flurry of media and public attention. In the press conference, David McKay, the leading author, adopted a cautious tone to explain why they considered "biologic activity" as the best explanation for the formation of the features identified in the meteorite [20]. The event got live coverage from the three main USA broadcast television channels and, in less than a week, NASA had counted more than one thousand stories about the research in USA television channels; this interest was also reflected in the extensive coverage of USA and international newspapers on the details of the research and its possible political consequences for the space agency [21]. Although it was later shown that the morphological structures in the meteorite were consistent with abiogenic processes [22] and the current consensus is that there's no compelling evidence of Martian microfossils in ALH84001, the public attention drawn by the claims generated political momentum for the institutionalization of astrobiology research at NASA [11, 12].

As research developed to include the search for possible alternative biochemistries, in 2011 the claim of discovery of "arsenic subsistent bacteria" in California's Mono Lake [23] was announced by NASA as an expansion of the scope of the SLE [24] and got ample international press coverage [25 - 28] A bitter public dispute among scientists in blogs and social media ensued, resulting in the refutation of the purported discovery [29]. In 2020, the announcement of the detection of a possible biosignature, phosphine, in the upper atmosphere of Venus [30], stirred the public sphere with a claim of possible life detection making the headlines in many international newspapers [31 - 36]. The detection was later refuted [37, 38], but it has been recently reappraised [39]. However, the biogenicity of the potential Venusian phosphine is still a contentious matter [40].

A suggestion of a possible detection of life outside of the Solar System appeared in September of 2023, with data from the James Webb Space Telescope (JWST) on the chemical composition of the atmosphere of the exoplanet K2-18b [41]. The paper claimed that some signals in the data collected from K2-18b's atmosphere could indicate the existence of a molecule called dimethyl sulfide (DMS), a known biosignature on Earth's environment. The statistical low confidence of detection of DMS ($1\sigma$) and the still insufficiently explored possibility of abiogenic processes giving rise to the molecule in a different and mostly unknown environment didn't stop the authors from speculating in the paper about the results amounting to "possible evidence of life" [41]. The next day, the headline on the BBC News website read: "Tantalising sign of possible life on faraway world" [42], whereas the author of the news story, science journalist Pallab Ghosh, described the result on social media as "new tantalizing evidence of life" on the exoplanet [43]. The story circulated widely in newspapers around the world, including in Brazil [44], Portugal [45] and the UK [46].

In that same month, BBC News followed up on the story with an article entitled "Alien life in Universe: Scientists say finding it is 'only a matter of time'", offering a very optimistic view of the possibilities of success of the SLE: “Many astronomers are no longer asking whether there is life elsewhere in the Universe. The question on their minds is instead: when will we find it? Many are optimistic of detecting life signs on a faraway world within our lifetimes - possibly in the next few years.” [47].

In similar tones, the expectation that the SLE will be successful in the near future has been expressed in the past decade by scientists, science communicators and journalists alike. Examples abound. In 2015, the American *National Public Radio (NPR)* quoted NASA leading scientists promising that “within a decade” there would be "strong indications" of “alien life” and "definite evidence" of it “within 20 to 30 years” [48]. In 2019, an article published by a science writer (an expert in a field unrelated to astrobiology) on the website *The Conversation* claimed that “discovery [of extraterrestrial life] now seems inevitable and possibly imminent” [49] and was widely republished by news outlets, including *Newsweek*, *The New Zealand Herald* and *Science Alert*. In 2021, *The Economist* ran a feature story entitled “The search for ET hots up: If life exists beyond Earth, science may find it soon” [50], offering an overview of current strategies of research in the SLE, optimistically describing the search for biosignatures in exoplanets and the promise future technology holds for the area. These are but a small fraction of optimistic news stories about the current state of the SLE.

Speculations, expectations and promises associated with the SLE are being communicated with the public and they play a role in the constitution of the public image of astrobiology. Understanding this portrayal should be part of an effort to establish strategies to adequately and effectively communicate research results from the field, especially those related to potential life detection. Since 2021, the scientific community has been attempting to establish a framework for assessing and communicating claims of biosignature detection [51, 52]. Related to these efforts, the Confidence of Life Detection (CoLD) numerical scale was proposed to guide the communication of research results [53]. It has since motivated at least one alternative proposal [54] and stirred debate [55]: for some in the community, adopting a numerical scale could backfire, fueling miscommunication [56] and even (unintentional) censorship [57]. It’s unknown whether the widespread adoption of a framework to assess and communicate research results would be detrimental or beneficial to the effective communication of the SLE.

With more data from exoplanets atmospheres coming from JWST (with which K2-18b's 2023 data was collected), the SLE has arguably entered a stage in which claims of possible life detection may reach the news cycle more quickly and in greater numbers than ever before. This context underlines the importance of understanding the prevalence of speculative content and expressions of expectations related to the SLE in the public sphere.

Studies focused on understanding how astrobiology is being portrayed in the public sphere are still scarce. Some touch on the subject incidentally. Mace & Schwalbe (2020) [58] have analyzed how American and British newspapers have framed the exploration of Mars from 2011 to 2016, a study in which the SLE is briefly approached as one of the frames identified by the researchers. They noted that the coverage of the *Curiosity* rover misrepresented the research goals of the mission, sensationalizing it as a search for life on Mars, which it wasn't [58]. News articles and press conferences were prominent primary sources used by Reinecke and Bimm (2022) [11] in a historical study on the shifts in strategies employed by astrobiologists to maintain justification for research related to the SLE in the face of failures in obtaining positive results, i.e. detecting life beyond Earth. Primarily aiming at characterizing the mechanism of justification of research in the face of lack of positive results called "the maintenance of ambiguity", this study was not focused on obtaining a clearer picture of the representations of the SLE in the media or the public sphere.

Directly touching on the question of the portrayal of the SLE in the public sphere, Schwarz & Seidl (2023) [59] conducted an analysis of the framing of astrobiology, the Search for Extraterrestrial Intelligence (SETI) and UAP (Unidentified Aerial Phenomena) in the German media. Their quantitative approach identified three main frames of what they call SETL (Search for Extraterrestrial Life) in the most widely read online news sources in Germany: the space exploration frame, the UAP (related to its extraterrestrial intelligence origin hypothesis), and the SETI frame. They found that scientific institutions are the most prominent sources of information for the German press in the space exploration and SETI frames [59].

The present study starts to fill that gap, shining light on the speculations, promises and expectations that surround the SLE and circulate in the science communication ecosystem. Three kinds of communications of scientific results are examined and compared: papers published in academic journals, press releases and institutional communication materials related to them, and newspapers articles reporting on the results of these studies. The timeframe of our study spans almost three decades: from August of 1996, when research claiming to have found clues of past Martian life in meteorite AHL84001 was published and received great public attention, until March of 2024. We examine the coverage of studies related to the SLE in six reference newspapers (*The New York Times*, *The Guardian*, *Folha de S. Paulo*, *Estadão*, *Público* and *El País*) in three different languages (English, Portuguese and Spanish) and from five different countries (United States of America, United Kingdom, Brazil, Portugal and Spain). During the timespan of our study, changes in the media landscape impacted news organizations' revenue generation, leading to layoffs and worse working conditions for journalists, with science coverage being hit especially hard [60]. Despite the decline of traditional media and the growing influence of digital platforms in news consumption, our choice to focus on newspapers stems from the evidence showing that traditional news organizations and legacy media are still the most trusted sources of information [61, 62].

Our goal was to understand how research results from astrobiology are being portrayed in the public sphere. We consider the study at least partially successful in this regard by uncovering the most frequent kinds of speculations and promises/expectations about the SLE being communicated with the public, and how they occurred in different kinds of documents.

## 1.2. Research questions

We sought to answer a broad, main question about the portrayal of astrobiology and the SLE in the public sphere by focusing on three specific sub-questions:

*RQ: How is research related to astrobiology and the SLE being portrayed in the public sphere?*

Specific sub-questions:

*SRQ1: What are the most frequent kinds of speculations and expectations/promises related to astrobiology and the SLE circulating in the media landscape?*

*SRQ2: How frequent are these speculations and expectations/promises in the scientific papers, press releases and news articles?*

*SRQ3: How frequent are these speculations and expectations/promises in different sub-areas of astrobiology?*

*SRQ4: In press releases and news articles, where are these speculations and expectations/promises occurring more frequently and to whom are they being attributed? (e.g. are they more frequent without attribution to experts or are they more frequent in quotes or attributed to experts?)*

# 2. Materials and Methods

This section presents the research questions, the development of the codebook and conceptual framework to answer those questions, the corpus and how it was built, and a brief report on the coding process and intercoder reliability of the instrument.

## 2.1. Corpus selection and organization

The English language corpus of news stories was composed by *The New York Times* (the leading newspaper in the United States of America in number of subscribers and digital readership) and *The Guardian* (a reference newspaper in the United Kingdom and without paywall). The Portuguese language corpus was constituted by news stories from three newspapers: *Folha de S. Paulo*, the leader in digital circulation in Brazil, *Estadão*, the Brazilian leader in print circulation [63] and *Público*, the leader in digital circulation in Portugal [64]. The Spanish language corpus is composed of news stories from *El País*, the leading newspaper in digital readership in Spain [65]. Alongside the language proficiency of two of the authors of the present study, the three languages were chosen because they are representative of a large proportion of the world population (they are among the top six in number of native speakers).

The search for the news stories were carried out using the built-in search mechanisms of each of the newspapers, except for El País, whose built-in search mechanism did not work with expressions in quotes. For this newspaper, the search was carried out in Nexis Uni [66]. Results found in Nexis Uni were then accessed in the newspaper website. This was in order to keep consistency in the access to the stories (all of the newspaper's stories were accessed, consulted and analyzed through their own websites). Keywords and expressions used to search for news stories had only slight variation in the three selected languages (see table 1).

**Table 1: Keywords and expressions utilized to search for news stories in three different languages**

| **English** | **Portuguese** | **Spanish** |
|---|---|---|
| Astrobiology; "life detection"; "evidence of life"; "evidence of biological"; "life elsewhere"; extraterrestrial; "extraterrestrial life"; "search for extraterrestrial" [life]; "search for life" and "searching for life"; biosignature; "habitable zone"; "habitable world" and "habitable worlds"; habitability; goldilocks; "goldilocks planet"; "alien life".<br><br>Keywords "alien" or "aliens" | *Astrobiologia; extraterrestre – extraterrestres; "vida extraterrestre"; "vida alienígena"; bioassinatura – bioassinaturas; "evidência de vida" e "evidências de vida"; "vida fora da Terra"; "busca de vida"; "zona habitável"; "planeta habitável".*<br><br>Keywords *alienígena*, *biomarcador* and the expression *"sinais de vida"* were dropped because they | *Astrobiología; extraterrestre* (filtering with Keywords: *planeta, vida, inteligente, microbio*); *"vida extraterrestre"; "zona habitable"; "planeta habitable"; biomarcador; biofirma; "vida fuera de la tierra"; "señal de vida"; "señales de vida" (extraterrestre); "búsqueda de vida"; "evidencia de vida".*<br><br>Keyword *extraterrestre* alone |
| returned too many results unrelated to astrobiology, like movie or TV show reviews. The word biomarker returned too many results related to health news. | returned too many unrelated results. | returned too many unrelated results, like movie and TV show reviews. |

Since we aimed to compare three different kinds of communication of scientific results, we built the corpus in *sets* containing at least one document of each kind: one news article, one corresponding press release and one corresponding paper. Some sets contain more than one document of each kind.

Our focus was to obtain news articles *reporting* on papers related to astrobiology and the SLE: because of that, we excluded editorials, book reviews, interviews, live coverages and stories unrelated to science coverage. News articles that were not related to astrobiology or the SLE were also excluded. We also excluded articles that did not have a study as a motivator or that did not cite or reference any studies. Only news articles reporting on published research studies (or studies that would be published), or that (directly or indirectly) referenced papers were selected. News articles that only cited or referenced conference communications but not published research papers were excluded.

News articles covering published research results from the Search for Extraterrestrial Intelligence (SETI) and other searches for technosignatures were included. Although there's an ongoing debate about the inclusion of SETI within astrobiology's multidisciplinary field, we follow here Erik Persson's conclusion that from a logical and philosophical perspective there is no question that SETI and search for technosignatures should be considered part of astrobiology [2]. Articles covering SETI initiatives are part of the corpus but are not prominent.

Once the news articles were found, we searched for the papers that were linked, cited or indirectly referenced by the news story. In the rare cases in which the paper was unavailable from the original source, the news article was not included in the corpus. With the corresponding papers found, we searched for the corresponding press releases or institutional communication. Most press releases were found through *Altmetrics*' compilation of stories and press releases citing the paper, or by directly searching on two websites specialized in distributing and replicating press releases, *EurekAlert!* [67] and *ScienceDaily* [68]. From the content provided by these platforms, it was possible to find the original press release in the research institutions websites. Some press releases were found through Google searches using the paper title, the authors' names and their respective scientific institutions. Given the comparative nature of our study, we also excluded news articles and papers for which a corresponding press release was not found or was non-existent.

A total of 272 news articles (118 in English, 118 in Portuguese and 36 in Spanish), 188 papers and 170 press releases (all of them in English) were selected for coding (in total, 630 documents). See S1 Appendix for a full description of the corpus organized in sets, with available links for each document (a substantial part of the corpus is behind paywalls). Fig 1 (below) shows the process of data collection.

**Fig 1. Data collection method.**

## 2.2. Conceptual framework and codebook

To identify the most common speculations, promises and expectations about astrobiology and the search for life elsewhere (SLE) circulating in the public sphere (see table 2), we developed a codebook specifically designed to compare three sources of information about scientific research: the original paper, the press release and the newspaper article. The definition of speculation used here is "the activity of guessing possible answers to a question without having enough information to be certain" [69]. The meaning of promise in this framework is related to expectation and optimism, "the idea that someone or something is likely to develop successfully and that people expect this to happen" [70]. Promises and expectations are interconnected: promising statements help create expectations of future benefits springing from scientific developments [71].

**Table 2. Types of speculations and promises/expectations related to the SLE, from our conceptual framework.**

| Speculations and promises/expectations | Category |
|---|---|
| **Speculations** | ***Outcomes***<br>Speculations about the *outcomes* of the SLE |
| | ***Significance***<br>Speculations about the *significance* of a particular research result for the SLE |
| | ***Evidence***<br>Speculations about the *evidential status* of a research result |
| | ***Existence***<br>Speculations about the *existence of life* beyond Earth |
| | ***Conditions***<br>Speculations about *conditions and/or ingredients for the existence/emergence of life* beyond Earth |
| **Promises/Expectations** | ***Detection***<br>The SLE is *expected to detect life* beyond Earth or *produce evidence* pointing to its existence |
| | ***Progress***<br>The SLE is making or will make *progress* |
| | ***Technology***<br>*Technological development* will/may provide clues, evidence or answers for the SLE |

The speculations were initially split into four different categories: about the *outcomes* of the SLE; about the *significance* of a particular research result for the SLE; about the *evidential status* of a research result; and about the *existence* of life beyond Earth. During the pilot study phase, it was clear a fifth category, about *conditions and/or ingredients* for the existence/emergence of life beyond Earth, should be added.

Among the promises and expectations, three kinds were distinguished: the SLE is *expected to detect life* beyond Earth or produce evidence pointing to its existence; the SLE is making or will make *progress*; and *technological development* will or may provide clues, evidence or answers for the SLE. These categories didn't change from the pilot studies to the actual coding phase, but for both speculations, promises and expectations, adjustments in the descriptions were made and clearer instructions and examples were added during the pilot study phase.

For each main category, if a speculation or promise/expectation was deemed to be present in the article, internal questions about the attribution of speculations and promises/expectations in the articles were applied. When speculations and promises/expectations occurred in press releases and news articles, we coded where they appeared and/or its attribution: in the title/subtitle, in the body of the text without being attributed to any source (i. e. it's expressed as part of the article written by the journalist or press officer and is not in quotes or attributed to an expert), in quotes of the authors of the study (or attributed to them), in quotes of an outside expert (or attributed to one of them), and with other attribution (e.g. attributed to an institution, the scientific community in general or to unidentified scientists/sources). We also coded the levels of speculative content for each category of speculation, but the coding of the levels was not sufficiently reliable between different coders to be included in the results (see section 2.3). The full version of the code book is available in S2 Appendix.

## 2.3. Intercoder reliability

To check the intercoder reliability of the coding scheme applied to the corpus in English, we randomly selected a sample of 14 news articles (11.8% of the total), 21 press releases (12.3%) and 21 papers (11.1%). After a review process and analysis of problematic cases of disagreements in the coding of the sample, with subsequent adjustments agreed upon by both coders, resulted in substantial agreement between coders, with an average of Krippendorff's alpha 0.819 for the main categories and 0.825 for internal questions about the attribution of the speculation and promises/expectations. Intercoder reliability was also checked for Portuguese and Spanish language corpora, but with a different second coder. A randomly selected sample of 12 news stories in Portuguese (10.1% of the total) and 5 news stories in Spanish (13.8%) was coded. After the same process of review applied to the English corpus, the average Krippendorff's alpha in the Portuguese/Spanish corpus for the main categories was 0.811 and 0.856 for the internal questions (about the attribution of speculations and promises/expectations in the articles). The coded levels of speculative content were not sufficiently reliable to be included in the results.

Full intercoder reliability results and description of the process of its development in pilot studies are available on S3 Appendix.

## 2.4. Data analysis

Since our research questions (1.2) aimed primarily at obtaining frequencies of occurrence of speculations and promises/expectations in different kinds of science communication articles, and where they occur inside press releases and news articles, we used descriptive statistics to present and analyze the results. Data analysis and visualizations were produced in Microsoft PowerBI, Microsoft Excel and Google Sheets from the data in the original Excel coding spreadsheet.

# 3. Results

## 3.1. The occurrence of speculations, promises and expectations

The five kinds of speculations occurred in papers, press releases and news articles at very different rates (see Fig 2). Speculations on *evidence* – about the evidential status of research results – were the rarest overall: they appeared in only 3.1% of papers, 3.5% of press releases and 5.8% of news articles. This is not surprising, given the extraordinary nature of the claims that underpin this kind of speculation: the research result may be or point to *evidence* of the (past or present) existence of extraterrestrial life. For example, claims derived from the original NASA study on possible microfossil structures in the meteorite ALH84001 appeared prominently in newspaper articles: "Evidence that a certain primitive form of life may have existed on Mars more than 3 billion years ago has been found in a meteorite that fell to Earth 13,000 years ago and was found in 1984" [72] (translated from Spanish); and: "scientists and space agency officials today reaffirmed their claim of finding strong evidence for past life on Mars and asked skeptics among the world's scientists to join them in conducting even more rigorous tests needed to confirm or disprove it" [73]. More indirect speculations are also examples of this kind: a press release claimed that the findings of a study about rocks in Mars' Nili Fossae region "may provide a link to evidence of living organisms on Mars, roughly 4 billion years ago in the Noachian period" [74].

**Fig 2: Speculations occurrence per article type.** Percentage of articles containing each type of speculation.

The speculation about the *outcomes* of the SLE was also very rare in papers (2.6%), albeit a little more frequent in press releases (7.6%) and news articles (11%). Examples of this kind of speculative content include simple and direct expressions like "scientists could even discover compelling evidence of aliens" [75], or expressions of optimism with the outcomes of the search in a timeframe, e.g. "it is quite possible that, within our lifetimes, atmospheric studies of these extrasolar planets will provide the first evidence of biological life beyond Earth" [76], or a perception of an increasing likelihood of success, e.g. "And with the discovery by Nasa and other scientists of Kepler 452b plus 12 other possible "habitable" exoplanets, it may have become just a little more likely that humans will find extraterrestrial life on another planet" [77]. More recently, a press release attributed to the authors of a study on a new hypothetical class of exoplanets a claim that their research results “could mean that finding biosignatures of life outside our Solar System within the next few years is a real possibility” [78].

Still rare in papers (5.3%), the speculation on the *significance* of research results for the SLE appeared in 28.2% of press releases and in 23.9% of news articles. These speculations deal with the meaning of the research results for the development and the goals of the SLE. For example, a 2021 press release from the European Southern Observatory portrays the result that points to the existence of an exoplanet with half the mass of Venus as “an *important step* in the *quest to find life* on Earth-sized planets outside the Solar System” [79] (our emphasis). In another example from ESO releases related to exoplanet detection (three around an M-dwarf star), the leading scientist claims the result is “a *paradigm shift* with regards to the planet population and the *path towards finding life* in the Universe” [80] (our emphasis). The fact they’re most occurring in press releases will be considered in the discussion section.

Appearing at a higher rate were the speculations on *existence* of life elsewhere. 14.3% of papers, 22.9% of press releases and 47.4% of news articles exhibited speculative content of this kind. Featured prominently in news articles, speculations about the existence of extraterrestrial life are frequently associated with the possibility of life on Mars. Examples of those include: “the finding raises the likelihood that any microbial life that arose on Mars may continue to eke out a rather bleak existence deep beneath the surface” [78] and “If it were possible to drill a mile into Mars into the newly discovered lake, he said he’d bet there was life there too” [79]. Speculations of this kind is also salient in earlier stories about studies on Martian meteorite ALH84001: ''This is supporting evidence for the presence of ancient life on Mars” and “the new study strongly supported the original claim and may even suggest that there is still microscopic life on Mars” [80] and, finally, "It means bacteria must have been very widespread on Mars" [81].

The most frequent kind, speculations on *conditions* and ingredients for life beyond Earth were present in 22.8% of papers, 47.6% of press releases and 56.6% of news articles. Examples of this kind of speculation linked to one study alone include direct claims, like “SwRI scientists discover evidence for a habitable region within Saturn's moon Enceladus" [82], and clear statements in newspapers headlines and fine line "Small Saturn moon has most of conditions needed to sustain life, Nasa says – Space organization finds that hydrogen erupts out of underground ocean on Enceladus, meaning it has the water, chemistry and energy sources life requires" [83]. In the same news article, scientists express optimistic claims about conditions and ingredients for life: "although we can't detect life, we've found that there's a food source there for it. It would be like a candy store for microbes" and "another scientist involved in the project said the discovery showed that the moon's ocean contained a potential chemical feast for microbes. "We have made the first calorie count on an alien ocean," he said" [86]. Another news story covering the same results channeled the scientist's optimism with favorable conditions for life in the Solar System: "With the first three of the four prerequisites ticked off, Coates now considers Enceladus, along with Jupiter's moon Europa, to be the most likely place in the solar system to discover microbial life today" [87].

The three kinds of expressions of promises and expectations also occurred in papers, press releases and news articles at different rates (see Fig 3). The rarest kind was *detection*, the promise/expectation that the SLE will detect extraterrestrial life at some point in the future, with only 0.5% of papers, 1.7% of press releases and 3.3% of news articles exhibiting it. A textbook example, the news article ran by *The Guardian* with the headline "'Mini-Neptunes' beyond solar system may soon yield signs of life" claimed that "signs of life beyond our solar system may be detectable within two to three years", attributing it to the authors of the study, one of them quoted: "we are saying that within two to three years we may see the first biosignature detection if these planets host life" [88].

**Fig 3: Promises/expectations occurrence per article type.** Percentage of articles containing each type of promises/expectations.

Promises/expectations that the SLE is making or will make *progress* occurred in 5.8% of papers, 22.9% in press releases and 20.5% in news articles. Contextualizing research results from an exoplanet characterization research, a press release illustrates this kind of promise/expectation very clearly: “This marks the first detection of an atmosphere around an Earth-like planet other than Earth itself, and thus is *a significant step on the path towards the detection of life* outside our Solar System” [89] (our emphasis). And the news article covering the same study followed through with the same approach: “The discovery marks one of the first times that scientists have spotted an atmosphere around a small, rocky world and brings them one step closer to the goal of finding life elsewhere in the universe.” [90].

The most frequent promise was the *technology* kind, expressing the expectation that technological development will or may provide clues, evidence or answers for the SLE, occurring in 11.7% of papers, 25.8% of press releases and 34.5% of news articles. In expressions of this kind of promise/expectation, new observation apparatuses like JWST feature prominently, like in a press release of the University of Cambridge: “These planets all orbit red dwarf stars between 35-150 light years away: close by astronomical standards. Already planned JWST observations of the most promising candidate, K2-18b, could lead to the detection of one or more biosignature molecules” [78]. The exoplanet K2-18b is also the protagonist of a news article containing this kind of promise/expectation:

> “Astronomers now hope to study more super-Earths for signs of water in their atmospheres. That work is due to be transformed in coming years with the launch of Nasa’s James Webb space telescope in 2021 and the European Space Agency’s Ariel mission in 2028. Observations from these telescopes should reveal more about the makeup of atmospheres on distant worlds, including the presence of methane and other gases that could direct signs of life.” [91]

With the exception of the *significance* speculation and the *progress* promise/expectation, speculations and promises/expectations exhibit a pattern of occurrence among the types of documents: papers have a significantly lower number of speculative content and expressions of promises/expectation, while press releases exhibit a higher occurrence rate than papers, and news articles are the type of document showing the highest occurring rate. In the *significance* speculation and the *progress* promise/expectation, however, press releases have slightly more occurrences than news articles, although papers continue to be, by far, the type of document with less occurrences.

## 3.2. The occurrence of speculations, promises and expectations in sub-areas

We also coded subareas of research related to astrobiology to detect any variations of occurrence of speculations and promises/expectations according to the general subject of study. Four broad subareas of research were distinguished: 1. *solar system* planetary science and robotic exploration, 2. *exoplanet research*, star-planet formation and interstellar medium, 3. *origins of life* and alternative biochemistries, and 4. Search for Extraterrestrial Intelligence (*SETI*) and technosignatures. A fifth category was added (*other*) to accommodate research that was not related to any of the aforementioned four.

Looking into the occurrence of speculations in each subarea of research in astrobiology, we found that 66.6% of articles from subarea *exoplanet research* exhibited at least one kind of speculation, with 60.5% in subarea *solar system*, 57.3% for *origins of life*. 66.6% of articles in subarea *SETI* exhibited at least one kind of speculation (see fig 4) – the same numbers apply for the category for *other*: 66.6%. Each sub-area had a different number of articles. The *exoplanet research* subarea had the most articles with speculation (186), and it also had the largest number of articles overall (279 in total).

**Fig 4: Percent occurrence of any speculation by sub-area.** In blue, the articles in which at least one speculation occurred; in grey, the articles in which no speculation occurred.

Promises/expectations also occurred most frequently in subarea *exoplanet research* (see Fig 5). 49.4% of articles from this sub-area exhibited at least one kind of promise/expectation, 27.4% in subarea *solar system*, 9.8% in subarea origins of life, 4.7% in subarea SETI, and 28.5% in the *other* category.

**Fig 5: Percent occurrence of any promise/expectation by sub-area.** In blue, the articles in which at least one promise/expectation occurred; in grey, the articles in which no promise/expectation occurred.

## 3.3. Attributions of speculations, promises and expectations in press releases and news articles

Tables 3 and 4 show percentages that correspond to the proportion of speculations and promises/expectations in each place/attribution in relation to the total of news articles or press releases that exhibited that kind of speculation or promise/expectation (papers weren't subject to these coding categories because they are, by definition, expressions of the authors). Note that the category "author" in these tables refer to the author(s) of the paper, not the author of the news articles or the press releases.

**Table 3. Percentage of occurrence of speculations by kind and by locus/attribution in news articles and press releases.**

| **News articles** | | | | | |
|---|---|---|---|---|---|
| | Title | Body | Author | Outside | Other |
| Outcomes | 3.3% (1) | 40% (12) | 33.3% (10) | 26.7% (8) | 10% (3) |
| Significance | 20% (13) | 43% (28) | 46.1% (30) | 21.5% (14) | 3% (2) |
| Evidence | 18.7% (3) | 43.7% (7) | 75% (12) | 25% (4) | 12.5% (2) |
| Existence | 34.1% (44) | 48% (62) | 45% (58) | 27.1% (35) | 10.8% (14) |
| Conditions | 37% (57) | 55.8% (86) | 53.9% (83) | 22.7% (35) | 7.8% (12) |
| | | | | | |
| **Press Releases** | | | | | |
| | Title | Body | Author | Outside | Other |
| Outcomes | 7.7% (1) | 46.1% (6) | 53.8% (7) | 0% | 7.7% (1) |
| Significance | 10.4% (5) | 43.7% (21) | 56.2% (27) | 20.8% (10) | 2% (1) |
| Evidence | 50% (3) | 83.3% (5) | 33.3% (2) | 16.7% (1) | 0% |
| Existence | 23% (9) | 43.6% (17) | 61.5% (24) | 7.7% (3) | 0% |
| Conditions | 17.2% (14) | 60.5% (49) | 64.2% (52) | 6.2% (5) | 1.2% (1) |

Total number of occurrences inside parenthesis. Color shades indicate the frequency of occurrence of speculations (darker shades indicate higher percentages). Column titles indicate the locus/attribution of occurrence. “Title”: the speculation occurred in the title and/or the subtitle of the article. “Body”: occurrence in the body of the text without being attributed to any source (i. e. it’s expressed as part of the article written by the journalist or press officer and is not in quotes or attributed to an expert). “Author”: in quotes of the authors of the study (or attributed to them). “Outside”: in quotes of an outside expert (or attributed to one of them). “Other”: with other attribution (attributed to an institution or to unidentified scientists/sources).

**Table 4. Percentage of occurrence of promises/expectations by kind and by locus/attribution in news articles and press releases.**

| **News Articles** | | | | | |
|---|---|---|---|---|---|
| | Title | Body | Author | Outside | Other |
| Detection | 11.1% (1) | 11.1% (1) | 66.7% (6) | 22.2% (2) | 11.1% (1) |
| Progress | 14.3% (8) | 46.4% (26) | 28.6% (16) | 35.7% (20) | 3.6% (2) |
| Technology | 0% | 59.6% (56) | 24.5% (23) | 26.6% (25) | 2.1% (2) |
| | | | | | |
| **Press Releases** | | | | | |
| | Title | Body | Author | Outside | Other |
| Detection | 0% | 33.3% (1) | 66.6% (2) | 0% | 0% |
| Progress | 5.1% (2) | 38.5% (15) | 59% (23) | 23% (9) | 2.6% (1) |
| Technology | 4.5% (2) | 56.8% (44) | 45.4% (20) | 15.9% (7) | 0% |

Total number of occurrences inside parenthesis. Color shades indicate the frequency of occurrence of promises/expectations (darker shades indicate higher percentages). Column titles indicate the locus/attribution of occurrence. “Title”: the promise/expectation occurred in the title and/or the subtitle of the article. “Body”:

occurrence in the body of the text without being attributed to any source (i. e. it's expressed as part of the article written by the journalist or press officer and is not in quotes or attributed to an expert). "Author": in quotes of the authors of the study (or attributed to them). "Outside": in quotes of an outside expert (or attributed to one of them). "Other": with other attribution (attributed to an institution or to unidentified scientists/sources).

In the tables, percentages of occurrence of each speculation and promise/expectation sum up more than 100% because the same speculation or promise could occur (and often did occur) in more than one place and/or with more than one attribution. For example, in the news story published by *The Guardian* in 2014 entitled "Ocean discovered on Enceladus may be best place to look for alien life" [92], speculations about the *existence* of life elsewhere occurred at the same time in the title/subtitle, in body of text with no attribution to experts, inside quotes of the author of the paper, inside quotes of an outside expert and attributed to undetermined scientists/experts. This is a very rare occurrence. However, it serves to illustrate that the same speculation or promise/expectation can be present in more than one place and/or with more than one attribution in news articles and press releases (e.g. it can be inside quotes of the author of the study *and* attributed to an outside expert in the same article).

When speculations occur in news articles and press releases, we have found that they appear more frequently in the headlines and subtitles of news articles than in titles of press releases – the only exception is for the *evidence* speculation, which appear in 50% of titles of press releases in which it occurs (albeit with a very low total number, 3), and in only 18.7% of headlines of news articles in which it occurs.

In press releases that presented speculations, the most frequent occurrence was inside quotes of the author(s) of the study (or attributed to them), with the exception of the *evidence* type: 53.8% for *outcomes*, 56.2% for *significance*, 61.5% for *existence* and 64.2% for *conditions* speculations. In comparison with press releases, news articles that contained speculations had them occurring inside quotes of the author(s) of the study in lower frequencies for four types: 26.7% for *outcomes*, 47.2% for *significance*, 35.6% for *existence* and 51.9% for *conditions*. The outlier was the *evidence* speculation: 75% of its occurrences in news articles had them attributed to an author of the study, and 33.3% in press releases (it should be noted the low total number of this occurrence, 2).

We also found that the occurrence of speculations associated with an outside expert in press releases were comparatively low: 0 for the *outcomes*, 7.7% for *existence*, 6.2% for *conditions* speculations and 16.7% for *evidence* speculations. In contrast, news articles exhibited speculations associated with outside experts more frequently: 26.7% of *outcomes*, 25%% of *evidence*, 27.1% of *existence* and 22.7% of *conditions* speculations were inside quotes of or attributed to an outside expert in news articles. The exception was the *significance* speculation, for which news articles and press releases presented similar percentages (21.5% and 20.8%, respectively).

With the exception of the *evidence* speculation, the second most prominent locus of occurring speculations in press releases was in the body of the article, not attributed to an expert or other sources. 46.1%% of *outcomes*, 43.7% of *significance*, 43.6% for *existence* and 60.5% of *existence* speculations appeared in the body of the text in press releases (the outlier was the *evidence* speculation, 83.3%). In news articles in which speculations occurred, 55.8% of *conditions*, 48% of existence, 43% of *significance* and 40% of *outcomes* speculations appeared in the body of the text without attribution to experts. These percentages are approximately at the same level of press releases. The only substantial discrepancy is in the *evidence* speculation: when it occurred in news articles, 43.7% of the time it was in the body of the text, while in press releases the percentage reached 83.3%, albeit with a very low total number (5). Finally, in news articles, we found a low occurrence of speculations in the “other” category – which is applicable to attributions to scientific institutions (e.g. NASA, ESA) or undetermined experts and sources, and an even lower frequency of this kind of attribution of speculations in the press releases (only 3 occurrences in total).

Promises/expectations appear in a very small fraction of the headlines and subtitles of news articles (14.3% for *progress* kind, 11.1% for *detection* and none for the *technology* kind), and of titles and subtitles of press releases (5.1% for *progress*, 4.5% for *technology* and none for *detection*). In press releases, 66.6% of *detection* (with a very low total number: 2 occurrences), 59% of *progress* and 45.4% of *technology* promises/expectations appeared in press releases inside quotes or attributed to the author(s) of the paper. In news articles, the *detection* promise/expectation also appeared prominently (66.7%) inside quotes or attributed to the author(s) of the study (although with a very low number, 6), while substantially lower (in comparison with press releases) for *progress* (28.6%) and *technology* (24.5%). In both news articles and press releases, the *technology* promise/expectation was most prominently present in the body of the text, in similar levels (59.6% and 56.8% respectively).

Among news articles in which promises/expectations occurred, 35.7% of *progress*, 26.6% of *technology* and 22.2% of *detection* were exhibited inside quotes or attributed to an outside expert. In this regard, this is consistently higher than press releases: 23% for *progress*, 15.9% for *technology* and none for *detection*. Both speculations and promises/expectations, then, appeared in lower proportion attributed to outside experts in press releases than in news articles.

Finally, as it happened with speculations, we found a very low occurrence of promises/expectations in the “other” category: five times in total for news articles (11.1% for *detection*, 3.6% of *progress* and 2.1% of *technology*) and only once (2.6% of *progress*) in press releases, which exhibited no promise/expectation of detection and technology in this manner.

# 4. Discussion

The significance of the SLE to the broader cultural and societal landscapes is usually portrayed as universal, articulated as a way of answering what is sometimes touted as the “greatest question of all” [93], a time immemorial riddle irrespective of cultural boundaries. The question is also portrayed as something that goes beyond science: finding life elsewhere, so it goes, may alter current assumptions about reality [94] and it’s recognized as having profound consequences of extra-scientific (philosophical, religious, ethical) character, potentially changing worldviews across different cultures [95]. Finding life beyond Earth is also viewed as the potential completion of the Copernican Revolution and Darwinian worldview: a cosmos in which the Earth and humanity have no privileged place [96].

Although there's no denying that the potential discovery of extraterrestrial life may have profound social and cultural impacts much beyond science, the grandeur of significance of the SLE can also be used as a rhetorical tool to foster and justify something more mundane: funding for research programs, with impact in the financial incentives for technoscientific projects and for the science communication ecosystem. Coupling this image of utmost significance of the SLE for humanity with the promise that it's reaching a point in which definite answers might be produced is a powerful way of setting up high public expectations about astrobiology and socially legitimizing it.

Expectations are decisive elements in the establishment of new scientific fields [97]. Promises of scientific progress and speculations about scientific breakthroughs play a role in fostering optimism with the potential development of a new area of research, constituting the horizon with which funding decisions are made, and the expectations they generate influence agenda setting [98]. Science hype, the raising of expectations around the prospects of success of a field, can be a positive force by attracting the attention and inviting the contribution of the public and a plurality of societal actors to help shape the future direction of research [99]. On the downside, failure of expectations can be damaging to the credibility of institutions associated with them [100].

In scientific discourse, overpromising (promises that are unwarranted by the current state of knowledge) is a tool for garnering funding and may lead to unrealistic expectations [101]. But distinguishing promising from overpromising can be contentious when the state of knowledge is still under development or under debate [101]. Speculative content, while not bound by strict demands of being accurate and truthful, influences societal factors involved in shaping the future of scientific and technological development [102].

The publication, in 2021, of a journalistic article asking "Why Do So Many Astronomy Discoveries Fail to Live Up to the Hype?" [103] inspired a synthetic analysis on the dynamics of hype in astronomy and astrobiology [104]. The journalistic article doesn't claim that there is an increase in hype in astronomy, just that there are many cases of hyped up results that didn't live up for the excitement created around them. Hype in science is usually blamed on the mass media, but it is probably a product of various entities involved in the production and circulation of knowledge [105, 106]. Lenardic et al, 2022 [104] looks at the possible dynamics of hype in astronomy and astrobiology and points to a systemic cause as the root of the phenomenon, with no distinguishable regulatory solutions like the establishment of frameworks for the communication of results in the SLE. It also takes for granted the *rise* in hype in astrobiology [104]. Although it is known that there's pressure on scientists to describe their research with overtly promotional words [107], as well as on science communicators to grab attention from media outlets and on journalists to write newsworthy stories [108], it is unclear if there is a *rise* in hype in astrobiology.

The speculations, promises and expectations that occurred in the articles analyzed in our research help shape the public image of astrobiology and the SLE. By playing this role, they can also influence the future development of the area by informing social and political processes that impact decisions about the direction of the research in the field. By circulating expectations and speculative information about the SLE in the public sphere, scientists, institutional science communicators and journalists are (intentionally or unintentionally) influencing and even shaping up the focus and the scope of current and future research. Expressions of optimism have the potential to lead to more funding in developing a particular technique or observational apparatus to seek for Earth-like planets (e.g. the *Habitable Worlds Observatory*), or opening up space for a particular field of research (e.g. exoplanet atmospheric characterization).

Since the conceptualization of this research, we have adopted a position of agnosticism towards life beyond Earth and the search for it. The current state of scientific knowledge about the prevalence of life in the universe is still highly uncertain. We don't know if life exists elsewhere in the universe and we don't know if we are ever going to reach a definite answer to the main question underlying the SLE. This discipline of thought and interpretation was needed to be able to spot the speculations, promises and expectations related to the field, since the current *zeitgeist* is dominated by the perspective that life (at least microorganisms) is probably common throughout the universe and it is conceivably within our grasp to find it elsewhere.

Speculations, expectations and promises are frequently latent, context-dependent content. The context for the occurrence and detection of these types of content is composed by the overall tone and approach of the article and the surrounding paragraphs, the state of scientific knowledge at the time of their occurrence, what the research result is telling and how information given about the result is being extrapolated in affirmations that are not supported by them. This is the relevance of coding the material in sets of associated papers, press releases and news articles. The approach gives the coder part of the necessary context to distinguish features of the communication of a research result such as the speculations and expectations/promises.

## 4.1. Comparing papers, press releases and news articles

In all coded categories the papers were, by a substantial margin, the type of document with least occurrence of speculations and promises/expectations. That's hardly surprising. Papers are the most rigorous and vetted form of communication of scientific results, subject to a publishing process that limits the space for speculations. Negative incentives are also at play: credibility might be damaged among peers if authors speculate – especially if the speculation is unwarranted by the results it describes or if it goes beyond current state of knowledge. We found that less than 6% of papers showed speculations, promises and expectations in five categories (*significance*, *outcomes* and *evidence* speculations; *detection* and *progress* promises/expectations). But papers also exhibited a relatively substantial amount of speculations about the *conditions* and ingredients for life (22.8%), about the *existence* of life (14.6%), and the expectation that *technological development* is going to provide clues or answers for the SLE in the future (11.7%). Consistently, in these categories press releases scored higher than papers (22.9%, 47.6%, 25.8% respectively) and news articles scored higher than press releases (47.4%, 56.6%, 34.5% respectively).

In only one kind of speculation (*significance*) and one kind of promise and expectation (*progress*) press releases were the type of article that exhibited proportionately the most occurrences. These are also the only categories that do not show the general pattern in which papers exhibited less speculations, promises and expectations than press releases, and press releases exhibited less speculations, promises and expectations than news articles. 22.9% of press releases, 20.5% of news articles and only 5.8% of papers had expressions that suggested that the SLE is making or will make *progress*. 28.2% of press releases, 23.9% of news articles and only 5.3% of papers had speculative content related to the *significance* of a particular research result for the SLE. This slight preponderance of speculations, promises and expectations of these kinds in press releases is probably related to the nature of both categories: they are associated to the communication of the relevance of a given research result to the progress of the SLE.

Science communicators responsible for press releases have been effectively acting as interpreters of published papers, often making narrative choices and deciding how research results are going to be framed [109]. The number of press releases produced and the number of them that reach the news coverage are the metrics typically used by PR Offices to assess their contributions to the success of their institutions' publicity [110]. Institutions employ PR to elevate their status in the public domain and among their peer competitors, aiming at increasing and justifying research funding. There's considerable motivation for the institutions to inflate the impact and/or the scope of the research results they are promoting into the news. This is why a major challenge for press information officers (PIOs) is "to resist the pressures of their top management" [111]. Our results show that they are the type of document that most circulates speculations about the *significance* of results and expectations of *progress* in the area, and this may be related to the way press releases are written to grab attention from the press by promoting the impact of the research results for the SLE.

The lower occurrence of speculations and promises/expectations attributed to outside experts in press releases in comparison with news articles is hardly surprising. Institutional communication materials like press releases usually have the authors of the study as authoritative sources and rarely have scientists not involved in the research to serve as a source of contextualization and explanation of the findings it describes. News articles, on the other hand, reflect a principle of science journalism: to seek sources not involved in research for feedback, context and possible skeptical considerations.

An interesting finding in both press releases and news articles is that the *detection* promise appeared much more prominently inside quotes of the author(s) of the study (or attributed to them), although the total number of occurrences of this type of promise/expectation was low overall in both types of documents (11 in news articles and 3 in press releases). That may indicate that when it comes to overt promises of life detection, journalists and science communicators tend to leave it in the hands of the scientists that made them, but the low total number of occurrences of the *detection* promise is a natural limitation for any interpretation about this.

Press releases and news stories, almost by definition, should not mimic the arid and controlled discourse of the paper, but, ideally, should still adhere to standards of rigor that limit expressions of speculative content.  When papers contextualize research results, it's usually directed to the peers, other experts from the area. That's part of the reason why they contain consistently less speculative content than press releases and news articles. A paper containing an announcement of the detection of a low-mass exoplanet in the habitable zone around a red dwarf star doesn't need to inform the involved community that the planet might offer conditions for the existence of life: it's implied. But the press release and the news article communicating the research result would miss the relevance of the study for the SLE if they didn't mention this possibility. It's in the public interest that contextualization of research results is communicated, especially when it articulates the results with the broader picture of a scientific field. The point is that there are many ways of contextualizing research results, exploring their significance and impact for the area, and pointing the way towards the future of the SLE. To speculate that a low-mass exoplanet around a red dwarf might be habitable is legitimate, but it would be a stretch to make optimistic statements about the likelihood of its habitability or of the existence of life in it before any further characterization of the planet, including its atmosphere.

## 4.2. Illustration of internal dynamics

Papers that speculate, even if cautiously so, can lead to less cautious speculative content in press releases and news articles. The communication of research results on the TRAPPIST-1 exoplanetary system is an example of a speculation present in the paper that got into the press release with a slightly different meaning, and finally made its way into a news article with that subtle, but noticeable, shift in meaning. We use here this set of documents as an illustration of how this dynamic can happen.

First, *conditions* speculations appeared in the paper:

> Using 1-dimensional and 3-dimensional climate models, Gillon et al. (2017) deduce that the surface temperatures of planets e, f, and g are suitable for harboring water oceans on their surface. However, using the runaway greenhouse limit of Pierrehumbert (2010), *we found that planet d, too, may be habitable,* if its albedo is $\gtrsim$ 0.3. *Planet d might be covered by a global water ocean that can provide a favorable environment for the appearance of life.* This conclusion supports the finding of Vinson & Hansen (2017), who calculate the surface temperature of the TRAPPIST-1 planets due to stellar irradiation and tidal heating due to the circularization of the orbits, *and found that planets d, e, and f might be habitable.* [112] (our markings).

The paper concluded that, *among the planets* in the system, "d and e are the most likely to be habitable." The press release, titled "TRAPPIST-1 System Planets Potentially Habitable", opened up with almost the same wording, but with a different meaning: "[t]wo exoplanets in the TRAPPIST-1 system have been identified as *most likely to be habitable*" [113] (our markings). Then, the headline of the article published by *The Guardian* on the same day of the press release read: "Two planets in an unusual star system are very likely habitable, scientists say" [114]. That's a stark shift in meaning: from an anodyne speculation in the paper to strong suggestions of habitable conditions in the press release and news article.

The TRAPPIST-1 system was the subject of another set of papers, press releases and news articles in which speculations and expectations abounded, albeit with a different dynamic. The paper announcing the discovery of the exoplanets didn't show any speculative content, nor promises and expectations. Two press releases were issued, one by the European Southern Observatory (ESO), the other by NASA. The carefully worded ESO article mentioned an increasing chance that the system "could play host to life" (*conditions* speculation) and quoted one of the authors saying that the then upcoming JWST would soon enable scientists to "search for water and perhaps even evidence of life on these worlds" (*technological development* promise) [115]. The NASA press release, also carefully worded, quoted one of the agency leading scientists stating the discovery "could be significant piece of the puzzle of finding habitable environments, places that are conducive to life" (*significance* and *conditions* speculations) and lauding the result as a "remarkable step forward toward [the] goal" of answering the question: "are we alone?" (*progress* promise) [116].

News articles followed. *The Guardian* published two stories on the discovery. The first article, compared to the follow up, was sober in tone but contained a speculation on the *significance* for the SLE in the headline "Discovery of new exoplanets is a lottery win for astronomers seeking alien life", alongside carefully worded promises of *progress* and *technological development* [117]. The second article reported on hopes being raised that "the hunt for alien life beyond the solar system could start much sooner than previously thought" (*progress* promise) "with the next generation of telescopes" (*technological development* promise), quoting one of the authors of the study saying "I think we've made a crucial step in finding out if there's life out there" (*outcomes* speculation) and that researchers hoped to know if there is life on those exoplanets "within a decade" (*detection* promise) [118]. *The New York Times*' article reproduced the same "crucial step" quote from one of the authors right after stating that "scientists could even discover compelling evidence of aliens" (*outcomes* speculation), juxtaposing this statement with another quote from the same scientist: "here, if life managed to thrive and releases gases similar to that we have on Earth, *then we will know*" (*detection* promise, our markings) [75]. This scientist is also quoted as saying that the research was "a step forward – a leap forward, in fact" towards answering the question that guides the SLE (*progress* promise) [75]. NYT also quoted an optimistic outside expert on how the discovery "make the search for life in the galaxy imminent" (*outcomes* speculation) [75]. Both NYT's article and *The Guardian* second story also contained *existence* and *conditions* speculations.

Although papers consistently exhibit less speculations and promises/expectations than press releases and news articles, some feature as prominent locus of occurrence of speculations and promises/expectations. The case of possible detection of a biosignature in the atmosphere of exoplanet K2-18b, from 2023, is illustrative. Except for the *outcomes* kind of speculations, the paper exhibited all of the other kinds of speculations and promises/expectations. The study speculates that a low confidence detection of DMS is a "possible evidence of life" on the exoplanet, and that the result "provides a pathway toward the possible detection of life on an exoplanet with JWST" [41]. Both press releases [119, 120] issued to promote the results didn't highlight the potential biosignature in the title or in the opening, and contained less speculations and promises than the paper (most occurrences in these documents inside quotes or attributed to the author(s) of the paper). Surprisingly, the news article [46] reproduced only one speculation, *conditions*.

## 4.3. Exoplanet research and the space for speculative content

The high occurrence of speculations and promises/expectations in articles related to exoplanet research is noteworthy. Even though this sub-area had roughly the same percentage of occurrence of speculations as the *solar system exploration* one (66.6% and 60.4%, respectively), it showed a higher propensity to exhibit promises/expectations (49.4% to 27.4%). A review article on the origins of life and the SLE associated the rapid growing number of detected exoplanets as a factor to speculate that "the chances of detecting signs of extraterrestrial life are increasing" [121]. Part of the motivation for research focused on detecting low mass exoplanets (presumably rocky worlds) in the habitable zone of their parent stars is attributed to the public interest or "public curiosity" about the theme [122]. As philosopher of science Peter Vickers stated once: "one thing unites [exoplanet scientists]: the search for extraterrestrial life" [123]. With the rapid growth of the number of detected exoplanets (from the first one in 1995 to 5000 by the end of 2022) and the current dawn of the era of exoplanet atmospheric characterization, it is hardly surprising that much speculative content occurred in articles from this subarea. Most importantly, with the general optimism with the prospects of detection of biosignatures in exoplanets' atmospheres in the near future, it's no surprise that expectations and promises were expressed substantially more in exoplanets research than in other subareas.

Speculations on *conditions* and *existence* of life in articles about exoplanet research are usually related to the habitability of planets outside the solar system and their possible similarity to Earth, which is usually conveyed by the term Earth-like. Potentially habitable exoplanets orbit their parent star in the so called "habitable zone", a distance to the star that may offer conditions for the planet to support liquid water on its surface (if other structural characteristics are present, like the presence of an atmosphere with the right pressure). One set of documents we analyzed is illustrative of how slippery it is to communicate about conditions for life beyond Earth when dealing with the concept of habitability. The paper dealt with a statistical estimate of planets in the "habitable zone" with data provided by the *Kepler* space telescope [124] and didn't show any speculations or promises/expectations, but the press release was titled: “Astronomers answer key question: How common are *habitable planets*?” and mentioned that some of them were subject to “lukewarm temperatures *suitable for life*” (our markings) [125]. On the same day of publication of the paper and the press release, *The New York Times* ran a story stating that “[t]he known odds of something — or someone — living far, far away from Earth improved beyond astronomers’ boldest dreams on Monday” and that the results “increased the chances that there might be life somewhere among the stars” [126]. *The Guardian* also published an article on that day with the headline: “Two billion planets in our galaxy may be *suitable for life*”. The story claimed that the data suggested “planets *capable of supporting life* are far more common than previously thought” and quoted one of the authors speculating that the universe seemed to produce “plentiful real estate for life that somehow resembles life on Earth” [127]. The drift in meaning from statistical inference of planets in the “habitable zone” to places “suitable for life” that may “resemble life on Earth” is remarkable.

The high occurrence of *conditions* and *existence* speculations may be related to the current state of knowledge about what is life, the limits of life and the necessary and sufficient conditions for its existence. Since knowledge about the existence of life beyond Earth and where the conditions for it might exist is still elusive, speculative content in the communication of the SLE seems to be almost inevitable: it reflects the still speculative and promising state of the young field. But the line that distinguishes unwarranted speculative content and legitimate contextualization seems very blurred.

Speculations and expressions of promises and expectations do not intrinsically constitute bad science communication or misrepresentations of the scientific process. In a field like astrobiology, especially in its SLE branch, speculative content is intrinsic in the development of concepts and hypotheses that guide the research. Nevertheless, not all speculations are legitimate in science, nor should speculative content be always prominent in a scientific field. It has been argued that the scientific process of testing hypotheses and winnowing out possible explanations with more and better empirical data should shrink the room for speculations in astrobiology [2]. Persson argues that room for speculations in the field should be limited by existing data and validated background scientific knowledge [2]. It's unreasonable to demand that the space for speculations and expectations in the *communication* of research results from the SLE should be limited in the same fashion of the scientific process. Although our research was not designed to identify a clear line to distinguish the space for justified speculative content from unwarranted speculations and overpromising, our findings do offer some useful clues for future efforts in making the communication of astrobiology and the SLE less subject to a kind of content that could artificially inflate public expectations and be harmful for astrobiology as whole in the long run.

## 4.4. Limitations and caveats

The research was not designed to capture speculations and expectations of the SLE in all the fields of research in astrobiology against the backdrop of all the news stories that were run about these subjects (e.g. exoplanets). The search was already designed to capture stories that contained mentions of the SLE or related topics (biosignatures, habitable zone, etc.). It is not an evaluation of the number of stories with some mention of the SLE against the backdrop of stories covering the same areas of research but without mentioning the SLE, the possibility of life, habitability and so on. To get over this limitation, it would be necessary to get all the news stories that were related to a particular subject (e.g. exoplanets, extremophiles) and quantify the stories that were framed as part of SLE against the backdrop of stories that weren't. Future studies should look into that.

The occurrence of speculations and expectations are not measured against the backdrop of all news stories related to astrobiology/SLE. It is only a comparative measure of the occurrence of speculations and expectations between papers, press releases and news stories that are related to a study and mention/allude to astrobiology and/or the search for life elsewhere. It is not a measure of how frequent are speculations in news stories about astrobiology in general (that would entail a different corpus, including stories that are not specifically related to a study). Further studies should focus on when there are skeptical evaluations or counter-speculations from the journalist, the same scientist or an outside expert.

This study analyzed documents in three languages, including articles published by selected newspapers in five countries. That's probably not representative of the whole world. Future studies should complement these results by looking into articles published in more languages by different newspapers from other countries. Also, since our focus was on legacy media, further studies should look into other media (blogs, social media and video platforms), which might present a different picture of the communication of astrobiology.

The coding framework has a limited ability to capture latent speculations, promises and expectations. Our chosen quantitative approach focuses on capturing explicit content, making implicit, more suggestive expressions of speculative and promising character intrinsically harder to isolate and pinpoint in the corpus. Future studies on the subject may use a qualitative approach to complement these results by capturing and characterizing latent speculations, promises and expectations in the same corpus here analyzed.

Potential influence of different funding agencies and scientific institutions in the communication of the SLE is a relevant topic not contemplated by our research design. We're looking to develop it in a future research project by adapting our framework to specifically identify instances of hype (unwarranted promotion of research) and spin, while also specifying a funding/institutional category and expanding the corpus.

While our results show that papers in general feature less speculative content compared to press releases and news articles, this does not warrant the inference that scientific authors or authors with a scientific background tend to be less speculative. Our study does not show that scientific authors tend to be less speculative, but that scientific articles are less speculative than press releases and news articles. We did not focus on the production process, but on the resulting texts. A relevant topic for future research is whether the scientific background (or lack thereof) of writers of press releases and news articles influences the article's tone and character.

# 5. Conclusion

This study reveals the complex landscape of the communication of astrobiology, highlighting the prevalence of speculations and promises/expectations in disseminating research findings. It underscores the role of news articles and press releases in shaping public perceptions of the SLE, often amplifying speculative content compared to academic papers. Although papers were, by a substantial margin, the type with least occurrence of speculations and promises/expectations, they still presented considerable speculative content in the *conditions* and *existence* categories. These were the most frequent speculations overall across all documents, highlighting how ample still is the current space for speculations in astrobiology's most elementary questions. Our results also point to news articles and press releases circulating in the public sphere the perception that the field is making progress towards finding answers and that technological development is going to provide clues or evidence for that. Coupled with room for speculative content, the optimism with advances in the near future of the SLE can boost public enthusiasm and influence funding decisions while drawing little attention to the current theoretical and empirical limitations of the field.

The potential for habitability and existence of life beyond the Solar System drives heightened promises and expectations in the communication of exoplanet research, which is also one of the coded subareas most susceptible to speculations. This study underscores the need for responsible communication practices that acknowledge the speculative dimension of astrobiology and balance optimistic views with realistic ones – this is especially important in communication of exoplanet research with regards to the prospects of confirming a possible biosignature detection. While public excitement and research funding are crucial for the development of the field, they should not be fostered through overly speculative and overpromising content. A key challenge is balancing the speculative nature of the field with informed decision-making and public understanding. This involves presenting research findings with appropriate caveats and clearly distinguishing between confirmed results and speculative interpretations.

Providing adequate context and accurate portrayals of research findings is essential. This includes explaining what is currently known and unknown about life (on Earth and elsewhere), the limitations of detection methods, the challenges of interpreting atmospheric data, and the uncertainties surrounding the concept of habitability. Scientists, science communicators and science journalists play a crucial role in ensuring this kind of responsible communication. They should be mindful of the potential impact of their words and strive to present a realistic view of the field, avoiding unwarranted hype or sensationalism.

By addressing these implications, the communication of astrobiology can be improved, fostering a more informed and engaged public while maintaining scientific integrity and credibility.

# Acknowledgments

The authors would like to thank the Lorentz Center staff for helping organize the Lorentz Center Workshop “Breaking News: We found extraterrestrial life”, which took place on 2-6 September 2024. We also thank all of the participants of the workshop for insights and feedback on partial results of this research presented during the event. We thank the researchers from the Earth-Life Science Institute in Tokyo for the feedback on partial results presentation during the visit of the corresponding author at the institution in May 2024. Vanessa Lourenço de Souza (masters student of the Science Communication masters program of the Campinas State University, Brazil) participated as the second coder for the Portuguese and Spanish corpora. Sanne Willems (assistant professor of Methodology & Statistics of the Institute of Psychology of the Leiden University, the Netherlands) greatly helped conceptualize data visualizations. Fabien Bijsterbosch (statistician, MSc in Statistics from Leiden University, the Netherlands) provided support in generating data visualizations.

# Supporting Information

**S1 Appendix. Corpus - Final Sets. Complete list of academic papers, press releases and news articles organized in sets**. (XLSX)

**S2 Appendix. Codebook. Comparative framework (paper, press release, news article): The portrayal of the search for life elsewhere in the media: speculations and expectations.** (DOCX)

**S3 Appendix. Full intercoder reliability results.** (DOCX)

**S4 Data set. The data set that supports the findings of this study containing the full coding spreadsheet is openly available in Zenodo at:** https://doi.org/10.5281/zenodo.15780329

**S5 Figure. Timeline of the contextualizing high profile claims in the search for life elsewhere with relevant events in astrobiology, including institutionalization and space exploration missions.** (TIFF)